\documentclass[%
 reprint,
superscriptaddress,
noeprint,
 amsmath,amssymb,longbibliography,
 aps,
prfluids,
onecolumn
]{revtex4-2}

\usepackage{graphicx}% Include figure files
\usepackage{dcolumn}% Align table columns on decimal point
\usepackage{bm}% bold math
\usepackage{ulem}
\usepackage{placeins}

\usepackage{caption}
\usepackage{subcaption}

\usepackage{lipsum}
\usepackage{color}
\usepackage{arydshln}

\begin{document}

\preprint{APS/123-QED}

\title{Making superhydrophobic splashes by surface cooling}% Force line breaks with \\

% in freezing spreading drop

%When growing ice crystals catch up an advancing contact line
%Ice crystals growth meets advancing contact lines on a freezing substrate

%\thanks{A footnote to the article title}%

\author{Rodolphe Grivet}
\email{rodolphe.grivet@ladhyx.polytechnique.fr}
\affiliation{Laboratoire d’Hydrodynamique (LadHyX), UMR 7646 CNRS-Ecole Polytechnique, IP Paris, 91128 Palaiseau, France}%

\author{Axel Huerre}%
 \affiliation{MSC,UMR 7057, CNRS-Université Paris Cité, 75013 Paris, France}%

\author{Thomas S\'eon}
\affiliation{Institut Jean Le Rond $\partial$’Alembert, UMR 7190, CNRS-Sorbonne Université,  75005 Paris, France}%

 \author{Christophe Josserand}
\affiliation{Laboratoire d’Hydrodynamique (LadHyX), UMR 7646 CNRS-Ecole Polytechnique, IP Paris, 91128 Palaiseau, France}%

\date{\today}% It is always \today, today,
             %  but any date may be explicitly specified

\begin{abstract}
%%ABSTRACT

We study experimentally the enhancement of splashing due to solidification. Investigating the impact of water drops on dry smooth surfaces, we show that the transition velocity to splash can be drastically reduced by cooling the surface below the liquid melting temperature. We find that at very low temperatures (below $-60 ^\circ \rm C$), the splashing behaviour becomes independent of surface undercooling and presents the same characteristics as on ambient temperature superhydrophobic surfaces. This resemblance arises from an increase of the dynamic advancing contact angle of the lamella with surface undercooling, going from the isothermal hydrophilic to the superhydrophobic behaviour. We propose that crystal formation can affect the dynamic contact angle of the lamella, which would explain this surprising transition. Finally, we show that the transition from hydrophilic to superydrophobic behaviour can also be characterized quantitatively on the dynamics of the ejecta.

%%ABSTRACT
\end{abstract}

\maketitle

%########## Intro ##########

\section{Introduction}

 Whether or not an impinging droplet splashes on a solid substrate is a widely studied question since several decades \cite{mundo1995droplet-wall}. It has been shown that inertia, viscosity and capillarity are key ingredients in determining the outcome of the impact. More recently, it was found that the surrounding air plays an important role on the splashing mechanism \cite{xu2005drop}, now understood as the source of aerodynamic lift and lubrication that levitate the fragmenting corolla. These findings were synthesised by Riboux and Gordillo in a model \cite{riboux2014experiments} taking into account the diversity of physical phenomena at stake in the triggering of splash. 

% Avec solidification

Initially motivated by its consequences in metallurgy, the sensibility of splashing to solidification has been first investigated twenty years ago, \cite{aziz2000impact,dhiman2005freezing-induced}, evidencing an enhancement of splash due to solidification, a phenomenon that was confirmed in recent times \cite{gielen2020solidification}. While several other studies focused on the solidification dynamics of drops impinging a cold substrate \cite{thievenaz2019solidification,thievenaz2020freezing-damped,kant2020fast-freezing,lolla2022arrested,dash2020computational}, the enhancement of splashing due to solidification is still puzzling and poorly characterised, although being critical for other applications, in particular for water drops, from spray coating to aircraft icing \cite{baumert2018experimental, cebeci2003aircraft,zheng2020inkjet}.

% Here we show
In this article, we thus experimentally investigate the outcome of millimetric water drop impacts on a dry smooth silicon substrate of varying temperature, from $100^\circ \rm{C}$ to $-140^\circ \rm{C}$. At positive temperatures, the threshold velocity above which droplets splash slightly increases with temperature due to continuous variation of air and liquid properties. For a substrate colder than the water melting temperature, this threshold strongly decreases with undercooling and reaches a minimum critical value under $-60^\circ \rm{C}$. We show that for colder substrates, the splashing is identical to the one on superhydrophobic substrates at ambient temperature. In particular, the threshold velocity, the ejected droplets angle and velocities are quantitatively the same. We explain this transition from hydrophilic to superhydrophobic impacts by an increase in the dynamic contact angle on cold surfaces, observed experimentally, and using existing splashing models. Finally, we propose a mechanism involving ice crystals formation modifying this dynamic contact angle of the spreading drop that would lead to this unexpected superhydrophobic behaviour.

\section{Experimental set up}

%####### FIGURE 1 ############
% Illustration de l'effet de la solidification avec 2 séquences, même vitesse et differentes T.
\begin{figure}
    \centering
    \includegraphics[width=\textwidth]{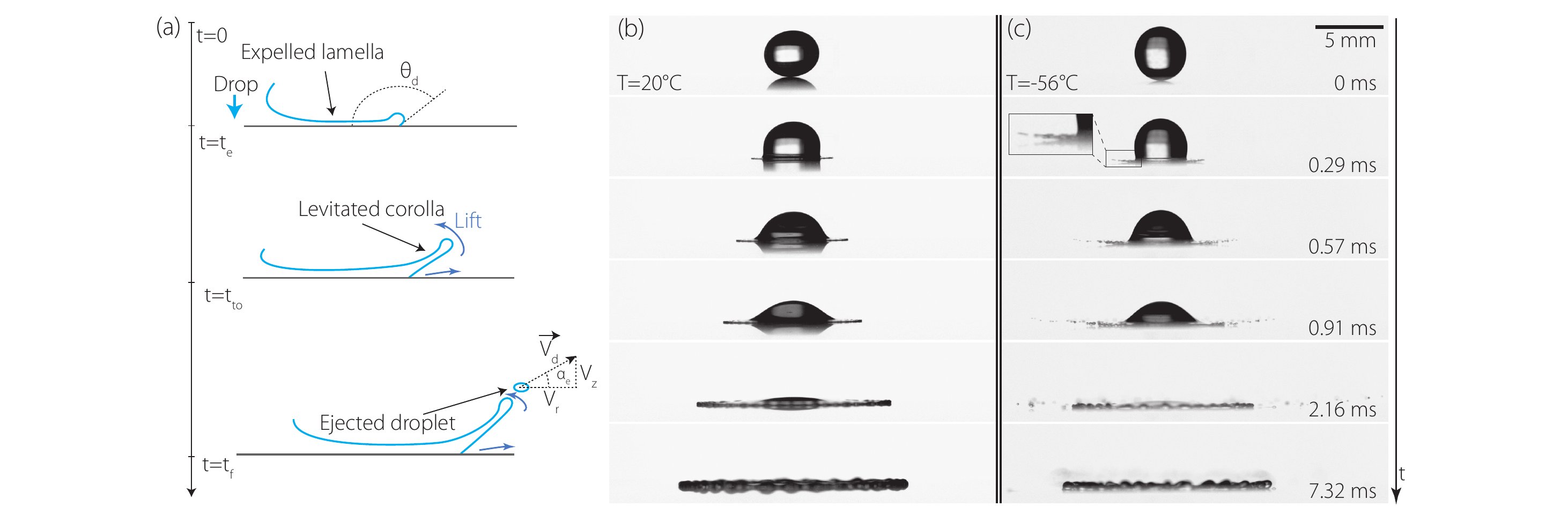}
    \caption{(a) Schematic sequence of a splash: impact and expulsion of a spreading lamella; lifting of the lamella to form a corolla; destabilisation of the corolla to form droplets. (b-c) Time sequence of an impact at various instants for $U_{\rm 0}= 2.65\, \mathrm{m.s^{-1}}$: (b) The surface is at room temperature, the droplet spreads, forming a thin lamella that expands. (c) The surface is at $\rm T=-56 ^\circ \rm{C}$, the lamella is lifted in the air to form a corolla, then destabilised in many droplets.}
    \label{fig:frise_t}
\end{figure}
%#############################

%############### Protocole XP ############
In the experiment, a pure water drop, at ambient temperature $T_{\rm w}=20^\circ \rm C$ and of constant radius $R=1.97  \pm 0.05\,\mathrm{mm}$ is quasi-statically released from a syringe at variable heights: $h=5 \:$-$\,140\, \mathrm{cm}$, leading to an impact velocity: $U_{\rm 0}= 1.15 \:$-$\: 4.7\,\mathrm{m.s}^{-1}$. The drop impacts a nanosmooth silicon wafer. The static (advancing) contact angle of the water drop is $\theta_{\rm s}=60° ^\circ$. The whole setup is placed inside a reduced humidity box %in order to avoid frost formation, 
and the relative humidity is set to $\mathrm{RH}=10 \%$ or $1 \%$ thanks to dryer modules and continuous nitrogen injection. No difference was observed between both humidity conditions. The impact is recorded from the side at 75.000 fps using a Phantom V2511 high-speed camera, with a spatial resolution of $30 \; \mathrm{\mu m} \cdot \mathrm{pixel^{-1}}$, and the drop is illuminated from behind. With these parameters, the drops Ohnesorge number is $Oh=\frac{\mu}{\sqrt{\rho R \sigma}}=2.6 \cdot 10^{-3}$, the Weber and Reynolds numbers are in the range $We=\frac{\rho R U_{\rm 0}^2}{\sigma} \in [35 \: ; \: 620]$ and $Re=\frac{\rho R U_{\rm 0}}{\mu} \in [2200 \: ; \: 9600]$, where $\mu = 10^{-3}\,\mathrm{Pa.s}$ is the water viscosity, $\rho=10^3\,\mathrm{kg.m^{-3}}$ is the density and $\sigma = 72 \, \mathrm{mN.m^{-1}}$ is the surface tension, taken at room temperature. The substrate temperature is varied between $+100^{\circ}\rm{C}$ and $-140^{\circ}\rm{C}$, using a heating plate for temperatures above room temperature, and by placing the substrate on an aluminium block cooled in liquid nitrogen for lower temperatures. The substrate temperature $T$ is measured at the time of impact using a thermocouple in contact with the surface, with a precision around $\pm 1 ^\circ \rm C$. After each impact, the wafer is cleaned by blowing dry air on it.

% Définition des termes

For each experiment, the outcome is determined between the two different possibilities: "Splashing", when at least one drop ejection is seen, and "Spreading" when no ejection is observed. In the following, as shown in Figure~\ref{fig:frise_t}(a) we will use the term "lamella" to describe the thin liquid sheet flowing along the substrate and fed by the impacting drop, and the term "corolla" to describe the lifted liquid sheet that necessarily precedes the splash \cite{latka2017thin-sheet}. It is commonly accepted that the splashing of a drop follows the usual sequence schematised on Figure \ref{fig:frise_t} (a): \textit{(i)} the drop impacts and a lamella is expelled at $t=t_{\rm e} $ \textit{(ii)} part of this lamella takes-off, hence forming a corolla at $t=t_{\rm to} $, and is then lifted upwards by aerodynamic effects, \textit{(iii)} this corolla destabilises into droplets \cite{xu2005drop,riboux2014experiments,riboux2015the-diameters}, ejecting the first one at $t=t_{\rm f} $.

%######## Description Figure 1 #############
\section{Results and discussion}

Figure \ref{fig:frise_t} (b-c) shows a time sequence of an impact at $\mathrm{U_{\rm 0}}=2.65 \, \rm{m.s^{-1}}$ on a wafer at ambient temperature and $-56 ^\circ \rm{C}$.
At room temperature (b), no splash is observed and the drop spreads following the classical sequence: impact ($\rm t=0\,$ms), creation and extension of a thin liquid lamella up to a maximum diameter ($0\,\rm{ms}<\rm t<7\,\rm{ms}$), and then retraction to the equilibrium spherical cap shape ($\rm t>7\,$ms, not shown) \cite{rioboo2002time,eggers2010drop}.
When the substrate is cold enough (c), the sequence is radically different. Soon after the impact ($\rm t=0.29\,\rm{ms}$) the liquid lamella has been lifted, forming a corolla in the air that has already destabilised, and we can observe droplets ejected on both sides (see magnification). As time goes ($0.29\,\rm{ms}<\rm t<1 \,\rm{ms}$) the size of the ejected droplets increases, a consequence of the increase of the corolla thickness \cite{riboux2015the-diameters,thoroddsen2012micro-splashing}. 
In this experiment, after about $  1\,\rm{ms}$, the corolla is totally fragmented and all the liquid flows in contact with the solid, following a classical spreading on a cold substrate \cite{thievenaz2020freezing-damped}.
From these two time-sequences, we can readily see that for a constant velocity the splash depends on the substrate temperature.
\\

%###### Figure 2 #################

%##################################

%# transition

\begin{figure}
    \centering
    \includegraphics[width=\textwidth]{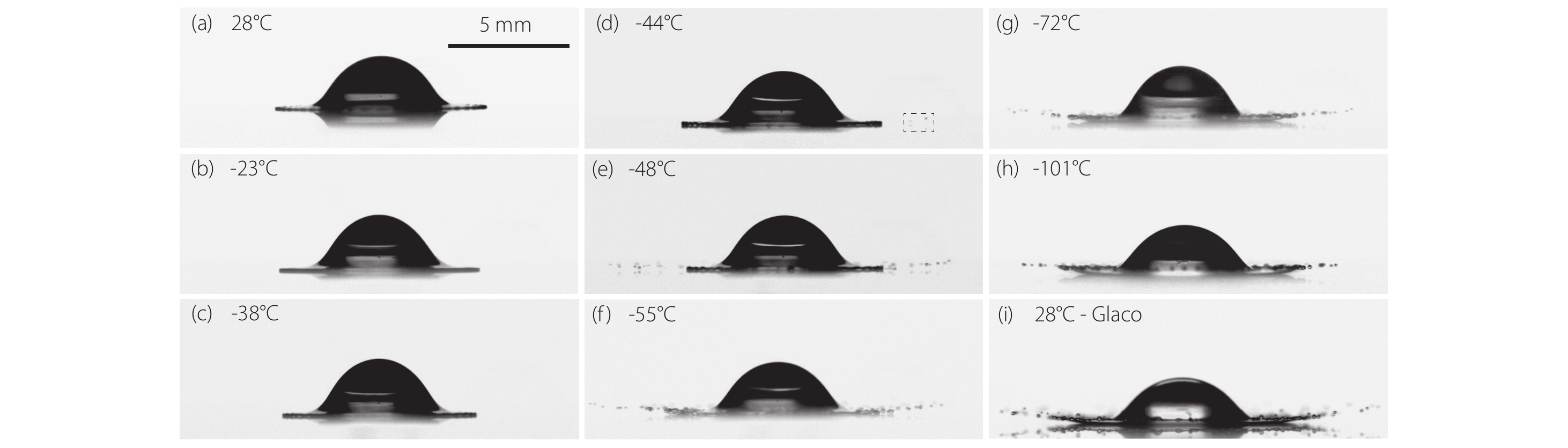}
    \caption{Impact outcome after $t=\tau_{\rm c}=R/U_{\rm{0}}$ for various temperatures indicated on each image at constant impact velocity $U_{\rm{0}} \approx 2.65 \, \rm{m.s^{-1}}$. (a-h) On an untreated silicon wafer, (i) on a Glaco covered aluminium plate. The first splash on the wafer is observed at $-44 {\rm ^\circ C}$ (dashed rectangle on (d)). }
    \label{fig:v_constant}
\end{figure}

%######### Description Figure 2 ########

To clarify the effect of the substrate temperature on the splash, we performed experiments at various substrate temperatures for $U_{\rm 0}=2.65 \pm 0.05 \, \rm{m.s^{-1}}$. We compare the results of the experiments at the characteristic impact time $\tau_{\rm c} =R/U_{\rm 0}$ ($0.72 \, \rm{ms}$ for each experiment) at which splashing is expected to have happened (see videos in Supplementary Materials). The corresponding snapshots are displayed in Figure \ref{fig:v_constant} (a-h) for temperatures between $28^\circ \rm C$ and $-101^\circ \rm C$.
We first notice that there is no clear difference between the droplet impacting at ambient temperature (a) and the ones at $\rm T = -23^\circ \rm{C}$ and $\rm T= -38^\circ \rm{C}$ (b-c). For the same time, $\tau_{\rm c}$, the spreading radius is almost the same, $r(\tau_{\rm c}) = 2.17R\: \pm 0.01 R$.
Cooling the substrate down to $-44^\circ \rm{C}$ (d) allows for the observation of ejected droplets on the right of the picture (see dashed rectangle).
The splashing transition temperature at this velocity is therefore located between $-38^\circ \rm{C}$ and $-44^\circ \rm{C}$.
Noticeably, for $T=-44^\circ $ and $-48 ^\circ \rm{C}$ (d-e), we do not observe a corolla but the ejected droplets exhibit the signature of early destabilisation. For colder substrates (f-h), the corolla is now clearly observed at $\tau_{\rm c}$ and never stuck again to the substrate.
Besides, the height reached by the first ejected droplet at $\tau_{\rm c}$ seems to increase with decreasing temperature, suggesting that the angle of droplets ejection ($\alpha_{\rm e}$, see Figure~\ref{fig:frise_t}(a)) also increases with undercooling, a fact that will later be related to the lifetime of the corolla.
Finally, it follows from the observation of the pictures that the number of ejected droplets increases strongly with the substrate undercooling, going from few isolated droplets ejected at $-44^\circ \rm{C}$ to continuous droplets ejection at lower temperatures.

%############## Figure 3 ##########################

\begin{figure}
    \centering
    \includegraphics[width=\textwidth]{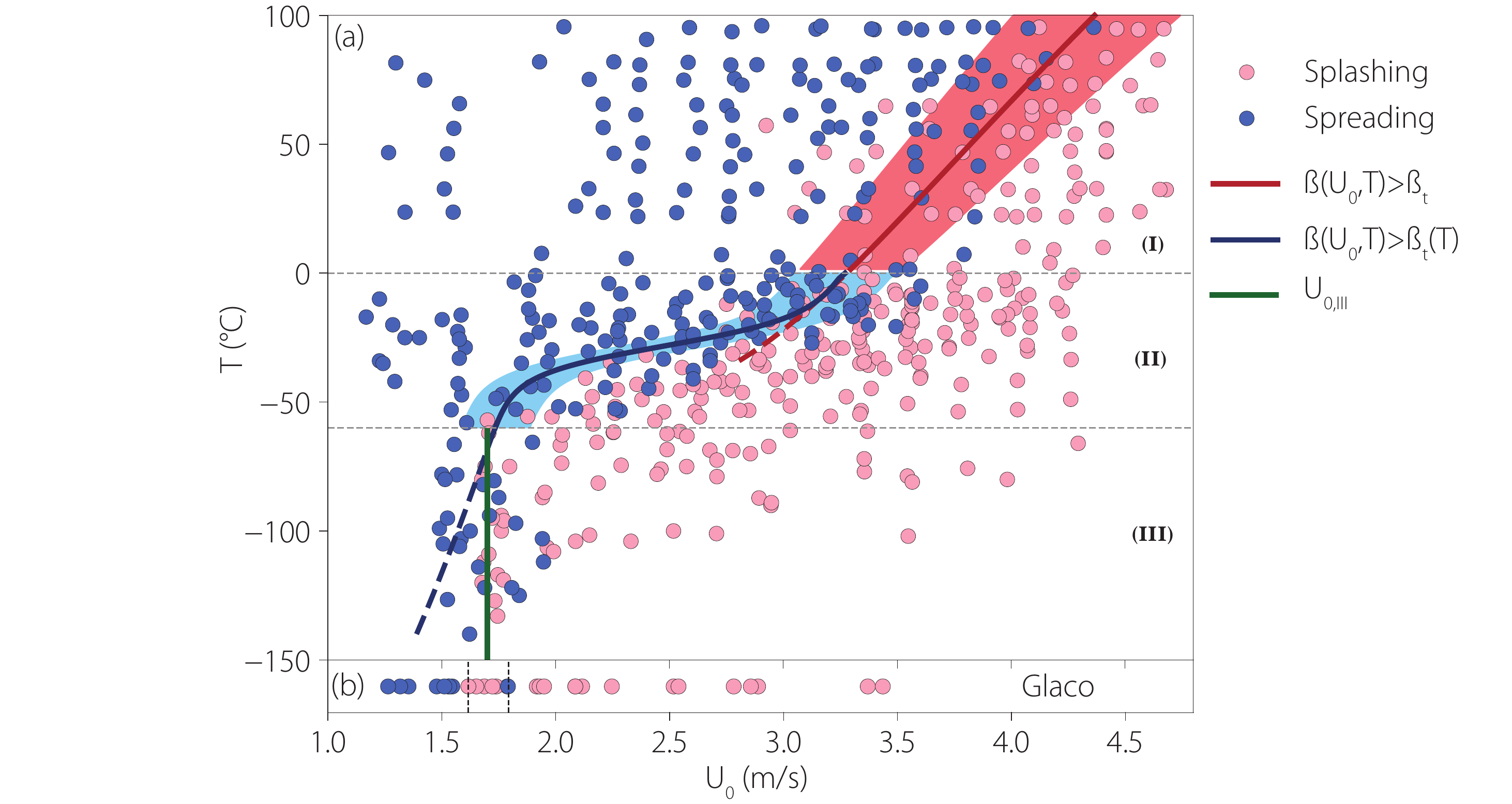}
    \caption{(a) Phases diagram of impact outcome as a function of substrate temperature and impact velocity. Two main domains of outcomes are distinguishable (Splashing and Spreading), and the transition velocity curve goes through three regimes depending on the surface temperature. The transition velocity in Region (III), $U_{\rm{0,III}}=1.7 \rm{m.s^{-1}}$, is represented by the vertical Green line. Red and Blue lines correspond to our modelling of the different Regions. (b) Impact outcomes on a superhydrophobic surface at ambient temperature. The transition velocity is similar to the one in Region (III).}
    \label{fig:phases}
\end{figure}
%###################################################

%######### Description Figure 3 ########
In order to characterise the coupled influence of surface temperature and impact velocity on splashing, Figure \ref{fig:phases} (a) shows the impact outcome as a function of these two variables. We see that two main domains are formed by the blue (spreading) and red (splashing) dots. The transition curve can itself be described in three different temperature regions, delimited by the horizontal grey dashed line.
%Region (I)
First, in positive temperatures (Region (I)), although the transition region is broad, the splashing velocity is observed to vary slowly with varying temperature, 
%no strong variation of the splashing velocity is observed with varying temperatures, and the transition to splash is set between $3.5$ and $4 \,{\rm m.s^{-1}}$ 
with only a 10\% variation over a range of $100^\circ \rm C$ for the upper bound.
% Region (II)
Secondly, between $0^\circ \rm{C}$ and $-60^\circ \rm{C}$ (Region (II)), a strong decrease of the transition velocity is observed from $3.5 \, \rm{m.s^{-1}}$ to $1.7 \, \rm{m.s^{-1}}$. 
% Region (III)
Finally, the transition velocity stabilises around a minimum below $-60 ^\circ \rm{C}$ (Region (III)), at $U_{\rm{0,III}} \approx 1.7 \, \rm{ m.s^{-1}}$, roughly half the one for the transition at room temperature, (green thick line in Figure \ref{fig:phases} (a)). Surprisingly, this minimum velocity is independent of temperature.

In the following, we seek to understand the physics behind the dependence of the splashing transition velocity on temperature.

%Discussion courbe (1)

\subsection{Region (I): A continuous variation of the physical properties with temperature}

 We first focus on Region (I), at positive temperatures, for which we adapt the isothermal splashing model developed by Riboux and Gordillo \cite{riboux2014experiments} in order to take into account the continuous variation of liquid and gas properties with surface temperature. The model focuses on the lifting criteria for the lamella, assuming that once a corolla appears, it will destabilise and produce a splash.
This is done by calculating the aerodynamic forces at the lamella expelling time, and by comparing them to the counteracting capillary one using the ratio of these forces, $\beta(R,U_{\rm{0}},\mu_{\rm w},\mu_{\rm g},\rho_{\rm w},\rho_{\rm g},\sigma,\lambda)$, where $\mu_{\rm i}$ are the liquid and gas viscosities, $\rho_{\rm i}$ are the densities, and $\lambda$ the mean free path of the air molecules. When $\beta>\beta_{\rm t}$, a threshold value, the drop is expected to splash. When the substrate temperature is varied, some of these physical properties become temperature dependant and it is important to determine which temperature should be considered for the different physical parameters of the model.

In fact, heat transfer during drop impacts has been recently analysed in a theoretical \cite{roisman2010fast} and experimental way \cite{moita2010heat} by considering how the advection of heat affects the purely diffusive behaviour in a one way coupling between the flow and the thermal problem. At high Prandtl numbers, $Pr_{\rm w}=\nu_{\rm w}/D_{\rm w}$, the thermal boundary layer is confined within the viscous one, which reduces the influence of the flow on the thermal problem. A self similar temperature profile establishes within the drop, and the substrate-liquid contact temperature is $T_{\rm c}=\frac{e_{\rm s} T+\mathcal{I}(Pr_{\rm w}) e_{\rm w} T_{\rm w}}{e_{\rm s}+\mathcal{I}(Pr_{\rm w})e_{\rm w}}$, where $e_{\rm s}$ and $e_{\rm w}$ are the thermal effusivities of silicium and water respectively ($e=\sqrt{\kappa \rho c_{\rm p}}$ for any material of thermal conductivity $\kappa$, density $\rho$ and thermal capacity per unit mass $c_{\rm p}$) while $\mathcal{I}$ needs to be integrated numerically \cite{roisman2010fast}. At ambient temperature for water, $Pr_{\rm w}=7.6$, which leads to $\mathcal{I} (Pr_{\rm w}) \approx 0.8$. Therefore, since $e_{\rm s} \approx 10 \,e_{\rm w}$, the correction due to advection on the contact temperature is negligible, on the order of $\approx 1^\circ \rm C$ at $T=-100^\circ \rm C$.

%The main idea of that model is to compute whether or not the expelled lamella can be elevated, assuming that once a corolla appears, it will destabilise and thus produce a splash. 
 With this contact temperature between the liquid and the substrate, we use the same approach as Staat \textit{et al.} \cite{staat2015phase}. To check that the current situation is the same as theirs, we similarly evaluate the ratio of the thickness of the thermal boundary layers in the liquid and in the gas ($\Delta_{\rm w}$ and $\Delta_{\rm g}$) to the size of the lamella ($H_{\rm l}$) at the time of lamella expulsion (or ejection in the terms of the original paper) $t_{\rm e}$, based on the characteristic residence time of liquid near the substrate and of air near the lamella:

\begin{equation*}
\frac{\Delta_{\rm w}}{H_{\rm l}} \propto \sqrt{\frac{\nu_{\rm w}(T)}{RU_{\rm 0} Pr_{\rm w} \frac{H_{\rm l}}{R}\frac{v_{\rm l}}{U_{\rm 0}}}} \approx  10^{-8/3}<<1
\quad \text{ and }\quad
\frac{\Delta_{\rm g}}{H_{\rm l}} \propto \sqrt{\frac{\nu_{\rm g}(T)}{RU_{\rm 0} Pr_{\rm g} \frac{H_{\rm l}}{R}\frac{v_{\rm l}}{U_{\rm 0}}}} \approx 10^{-1/3} \approx 1
\end{equation*}
where $\nu_{\rm i}$ are the kinematic viscosities, and $v_{\rm l}$ is the lamella velocity at $t_{\rm e}$. The low $Oh$ approximation has been made and prefactors have been dropped, allowing to write $\frac{H_{\rm l}}{R}\frac{v_{\rm l}}{U_{\rm 0}} \propto t_{\rm e}/\tau_{\rm c} \propto We^{-2/3} $.

In this situation, following Staat \textit{et al.} \cite{staat2015phase}, the lamella should be considered at drop temperature and thus the surface tension kept constant, whereas the viscosity is considered at $T_{\rm c}$.
Similarly, the gas is thermalised by the surface, hence at $T$, and so its density and mean free path vary with $T$, whereas its viscosity is kept at drop (ambient) temperature. 
The air density is 
$\rho_{\rm g}(T)=~10^5/(287\times T)\;\rm{kg\cdot m^{-3}}$ with $T$ in Kelvin,
the air mean free path is 
$\lambda(T)=68\cdot10^{-9}\times T/293\;\rm m$
and the supercooled water viscosity is taken from Dehaoui \textit{et al.} \cite{dehaoui2015viscosity} as $\mu_{\rm w}(T_{\rm{c}})=\mu_{\rm 0}(T_{\rm{c}}/T^\circ-1)^{-\gamma}$
where $\mu_{\rm 0}=1.3788\cdot 10^{-4}\;\rm{Pa\cdot s}$, $T^\circ =225.7 \; \rm{K}$ and $\gamma=1.6438$, valid down to $-34^\circ \rm C$. 
With these properties, we can then use the splashing model \cite{riboux2014experiments} with $\beta_{\rm t} = 0.13$ in order to reproduce best the trend of our data in Region (I). This value is not far from $0.14$ proposed by the authors.
The prediction given by this modelling is represented by the thick red line on Figure \ref{fig:phases} (a), where it can be seen that the decrease in aerodynamic forces due to heating creates in turn an increase of the impinging velocity necessary to splash. Left and right limits of the shaded areas show the predictions for $\beta_{\rm t}=0.125$ and $\beta_{\rm t}=0.135$ to illustrate the sensitivity of the model. 
We conclude that the small variation of the transition velocity in Region (I) is in accordance with this classical splashing model extended to temperature dependent variables.

% Justification solidification
The red dashed line in Region (II) on Figure \ref{fig:phases} (a) shows the predicted transition curve at negative temperatures, down to $-34^\circ \rm C$. Although it seems to reproduce well the trend observed at moderate supercoolings, as recently proposed using similar arguments \cite{lei2023splashing}, this framework can not be extended at lower temperatures. 
Below $T_{\rm f} \approx -40^\circ \rm C$, liquid water reaches the so called "No Man's Land", under which metastable liquid water properties where never measured \cite{koop2004homogeneous}. This is due to the high probability of ice nucleation, at these temperatures. Discarding any effect of the solid surface and only considering homogeneous bulk nucleation, experimental data show that the nucleation rate increases by a factor greater than $10^{18}$ over only $10 \rm K$ near $T_{\rm f}$ \cite{koop2004homogeneous,pruppacher1995a-new-look}. Therefore, extending the adapted splashing model to water supercooled below $T_{\rm f}$ would have no physical meaning, as the flowing water within the thermal boundary layer should start freezing immediately after contact. This vision is reinforced by the strong slope breakup close to the water melting temperature, which suggests that solidification is responsible for this abrupt change. Note that the presence of solidification implies that the flowing water in Regions (II) and (III) is at its melting point, hence $0^\circ \rm C$, a fact that will be used in the following.

\subsection{Region (II): Effect of solidification on the dynamic advancing contact angle}

\begin{figure}
    \centering
    \includegraphics[width=\textwidth]{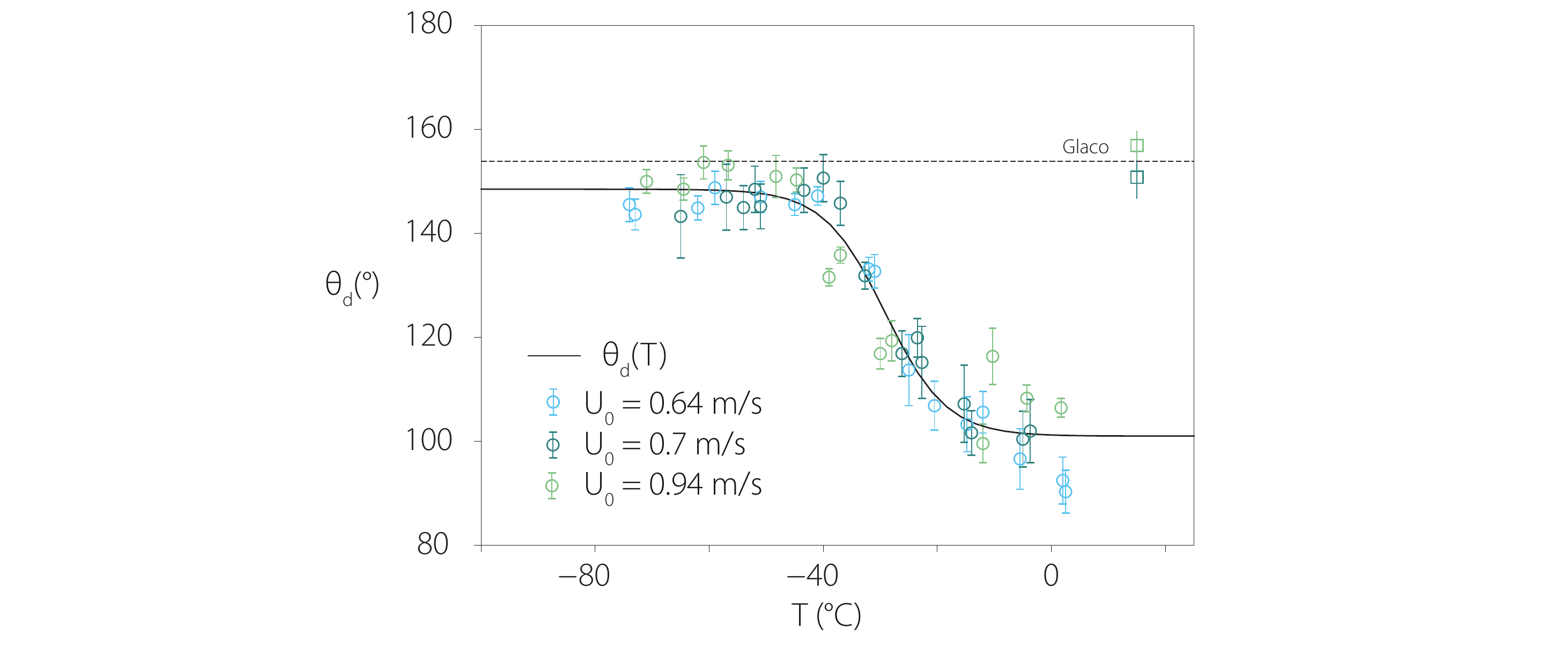}
    \caption{Dynamic contact angle of the lamella during spreading as a function of temperature for three different impact velocities. The solid black line is a sigmoïd fit of the points. The square symbols represent ambient temperature impacts on Glaco, and the dashed horizontal line the average value of these experiments.}
    \label{fig:angles}
\end{figure}

We now move to the understanding of the strong decrease of the splashing velocity that takes place in Region (II). In different experimental conditions, such drastic reduction of the splashing transition velocity without changing the liquid has already been observed in the well documented case of rough surfaces \cite{de-goede2018effect,garcia-geijo2021spreading,rioboo2001outcomes,aboud2015splashing,rioboo2002time} and of superhydrophobic impacts \cite{almohammadi2019droplet, aboud2015splashing,quetzeri-santiago2019the-effect,quintero2019splashing,zhang2021effect}. 
Insights on the effect of wettability on splashing were recently gained \cite{quetzeri-santiago2019role} by taking the perspective of lamella spreading. 
The authors found that the dynamic contact angle of the lamella during spreading, $\theta_{\rm{d}}$,  depends on the static (advancing) contact angle of the liquid-solid system, and that it determines the splashing of the drop. They showed that increasing this angle leads to a smaller splashing threshold: the hydrophobicity promotes the splash through the maximal dynamical contact angle. In fact they proposed that the threshold force balance $\beta_{\rm t}$ is a linear function of that angle, such that for a given impact, 

\[\beta_{\rm t}=-\alpha \theta_{\rm d} + \beta_{\rm 0}\]

where $\beta_{\rm 0}$ is the threshold value for a (theoretical) $0^\circ$ angle during spreading, and $\alpha$ is an empirical coefficient.

% Présentation des manipes
We performed a separate set of experiments in order to measure the dynamic contact angle reached by the lamella for various substrate temperatures during the spreading at moderate impact velocities, $U_{\rm 0} \in \{0.65, 0.71, 0.95\} \: \rm{m.s^{-1}}$, where no splashing could be observed, and hence where a dynamic contact angle could be measured. The experimental set-up is close to the one already presented, except that the spatial resolution is increased, around $3 \: \rm{\mu m / px}$ using a Navitar X12 lens, and that the temporal resolution is in turn decreased, around $35.000$ fps. These experiments were also carried out on aluminium plates covered with commercial Glaco coating, which present a $155^\circ$ equilibrium advancing contact angle with water (measured with sessile drop method). Previous studies on Glaco coated surfaces noted a very low surface roughness \cite{quetzeri-santiago2019the-effect,langley2018the-air-entrapment}, less than a micrometer in RMS, which should not affect the spreading nor the splashing behaviour by itself: the roughness-based Ohnesorge number being $Oh_{\rm r}=\mu/\sqrt{\rho \sigma R_{\rm{RMS}}} \approx 1 $, a value sufficiently large so that roughness should have no influence on splashing, as recently shown \cite{de-goede2021droplet}. The value of the dynamic advancing contact angle was obtained following the method proposed by Quetzeri-Santiago \textit{et al.} \cite{quetzeri-santiago2020on-the-analysis} (see Appendix \ref{appAngle} for further details). 

Figure \ref{fig:angles} shows the evolution of the apparent dynamic contact angle $\theta_{\rm d}$ as a function of surface temperature for the three different impact velocities investigated. Surprisingly, $\theta_{\rm d}$ increases with undercooling, from approximately $100^\circ$ to $150^\circ$ between $0^\circ \rm C$ and $-50^\circ \rm C$, with no visible dependence on the droplet impact velocity in the range studied. Note that the dynamic angle measured at positive temperatures, around $100^\circ$, is coherent with the one expected given the value of the advancing equilibrium contact angle of water on the silicon wafer ($60^\circ$) \cite{quetzeri-santiago2019the-effect}. The square symbols correspond to impacts on a Glaco coated surface at ambient temperature, and the dashed horizontal line represent the averaged value of these experiments, showing a dynamic contact angle around $155^\circ$ which is exactly the advancing equilibrium contact angle of water on Glaco, a result also in agreement with previous observations that on superhydrophobic surfaces, $\theta_{\rm{d}}$ has the same value as the  advancing equilibrium contact angle \cite{quetzeri-santiago2019the-effect}. The evolution of this dynamic angle with temperature is well fitted by a sigmoïd function $\theta_{\rm d}=f(T_{\rm s})$, shown in black on Figure \ref{fig:angles} (b), whose asymptotic values are $\theta_{\rm H}+\Delta \theta=148^\circ$ at low temperatures, and $\theta_{\rm H}=101^\circ$ at positive ones.
We can combine that relation with the expression for $\beta_{\rm t}(\theta_{\rm d})$, so that $\beta_{\rm t}$ is now a function of the surface temperature:

\[\beta_{\rm t} (T)=-\alpha  \Big [\frac{\Delta \theta}{1 + e^{-a(T - T_{\rm i})}} +  \theta_{\rm H} \Big ] +\beta_{\rm 0} \]

where $\theta_{\rm H}=101^\circ$, $\Delta\theta=47^\circ$, $T_{\rm i}=-29^\circ \rm C$ and $a=-0.19$ are the parameters of the sigmoïd fit.

Assuming that solidification brings the flowing water temperature back to $0^\circ \rm C$, whereas the air properties follow the same dependence on temperature as before, we can use our previously adapted model to predict the splashing transition velocity on surfaces at negative temperatures. We only accommodate $\alpha=0.007$, not far from $\alpha=0.0011$ proposed previously \cite{quetzeri-santiago2019the-effect}, which necessarily provides the value $\beta_{\rm 0}=0.20$ in order to match our data of $\beta_{\rm t}=0.13$ when $\theta_{\rm d}=101^\circ$ at positive temperatures. Note that this value of $\alpha$ would also be able to separate the original data of Quetzeri-Santiago \textit{et al.} \cite{quetzeri-santiago2019the-effect} due to their dispersion. This model is shown by the dark blue line of Figure \ref{fig:phases} (a), which accurately separates the splashing and spreading regimes in Region (II). The dark blue dashed line, which corresponds to the same model extended at lower temperatures, still shows dependence on surface undercooling in Region (III) due to the increase in aerodynamic forces as a result of the air cooling, and hence is not able to correctly predict the transition in that temperature range.

\subsection{Region (III): Superhydrophobic behaviour}

%Hydrophobe
 In the following, we finally wish to understand this intriguing temperature-independant minimum splashing velocity of Region (III), $U_{\rm 0,III}$, by making use of our previous observation that the impacts become closer to impacts on superhydrophobic surfaces as surface temperature is decreased.
Experiments on superhydrophobic substrates were hence also carried out in the original experimental set-up, in order to compare the effect of solidification on splashing transition velocity with the one of superhydrophobicity.

The results of these experiments are presented on Figure \ref{fig:phases} (b) together with a view of an impact at $t=\tau_{\rm c}$ at room temperature and in the splashing regime in Figure \ref{fig:v_constant} (i). 
Noticeably, the splashing threshold velocity on Glaco at room temperature is close to $U_{\rm{0,III}}$. This similarity between the two transition velocities is supported by the aspect of the splash, which strongly resembles the very low temperature splashes on the silicon wafer (Figure \ref{fig:v_constant} (g-h) ). 
Remarkably, cooling superhydrophobic surfaces from ambient down to $-70 \rm{^\circ C}$ has no effect on the splashing transition (see Appendix \ref{appSH}) suggesting that this minimum threshold transition velocity is already reached on superhydrophobic substrates. This also confirms that the un-adapted splashing model \cite{riboux2014experiments} should not be extended to negative temperature, as it would predict a reduction of the transition velocity with increasing undercooling also for the superhydrophobic substrates. In fact, it was recently shown \cite{quintero2019splashing} that the criterion on lamella elevation is not the relevant one in the case of splashing on superhydrophobic surfaces, the limiting process in that case being whether or not the rim of the flying corolla can destabilise by a competition between inertia and capillary forces. In that framework, solidification can not affect the splashing transition velocity, as the destabilisation mechanism only depends on the flow properties and on the liquid viscosity and surface energy in the rim, none of these being dependent on the surface temperature. Subsequently, $U_{\rm{0,III}}$ appears as a lower bound for the splashing transition velocity on any smooth surface, which explains the saturation phenomenon in Region (III).

In order to understand the value of that minimum velocity, we can compute the theoretical superhydrophobic dynamic angle that would lead to this threshold velocity using the dependence of $\beta_{\rm t}$ on $\theta_{\rm d}$. This leads to $\theta_{\rm d,SH}=159^\circ$, very close to the one measured in our experiments, $\theta_{\rm d,SH}=155^\circ$. Note that using this measured angle would lead to a theoretical transition velocity of $1.8 \:  \rm{m.s^{-1}}$, also close to $U_{\rm{0,III}}$.\\

%Courbe 2

%Intro Mexicains

Regarding the solidification mechanism behind our observations, we propose that the observed change in $\theta_{\rm d}$ can be induced by the presence of ice nuclei close to the advancing contact line. Indeed, Kant \textit{et al.} \cite{kant2020fast-freezing} recently proved in experiments with hexadecane that solid nuclei formed at the cold substrate surface are advected close to the contact line by the flow. A possible scenario would be that these advected crystals can reach the fast moving contact line, hence acting as defects and holding it, in a way similar to what has already been observed recently for the case of droplets spreading without inertia \cite{grivet2022contact}. These defects could induce an increase in the apparent spreading contact angle \cite{joanny1984a-model}, which would in turn explain the transition from hydrophilic to superhydrophobic behaviour.
A second possibility could be that the crystals, advected to the  water-air interface, would create a liquid-marble-like structure at the advancing edge of the lamella \cite{aussillous2001liquid}. This effect of ice crystals has already been experimentally established on drop impacts of water already partially frozen \cite{kant2020pattern}, and would lead to an increasingly hydrophobic-like behaviour with colder surfaces. 
Investigating experimentally these scenarios is highly challenging and would be worth a dedicated study.

At this stage, we have shown that in positive temperatures, the splashing transition is well modelled by a continuous variation of the air and liquid physical properties (Region (I)). Cooling further the substrate should induce ice nucleation, which has been experimentally shown to increase the apparent dynamic advancing angle of the lamella during spreading, which in turn lowers the splashing threshold  (Region (II)). When this threshold velocity reaches the one for superhydrophobic surfaces, the limiting mechanism for splashing is the corolla destabilisation, independent of temperature, and the splashing velocity settles at this unique lower bound value (Region (III)). 

%############## Figure 5 ####################

%###########################################

% ########## Description Figure 4 ##########

% Une fois très froid, même géométrie du splash quelque soit T
% Une fois très froid,même variation que super-hydrophobe quand U_{\rm 0} varie
% En dessous de -60, le splash ne change pas et a toutes les carac du superhydrophobe
% L'angle d'éjection varie clairement avec la température

%Insert : le mécanisme de déstabilisation est toujours le même
% Vs est proportionnelle à U_{\rm 0} pour toutes les T et tous les substrats
%

%Figure principale
%Transition via la morphologie

\section{Effect of solidification on the ejecta dynamics }

\begin{figure}[h!]
    \centering
    \includegraphics[width=\textwidth]{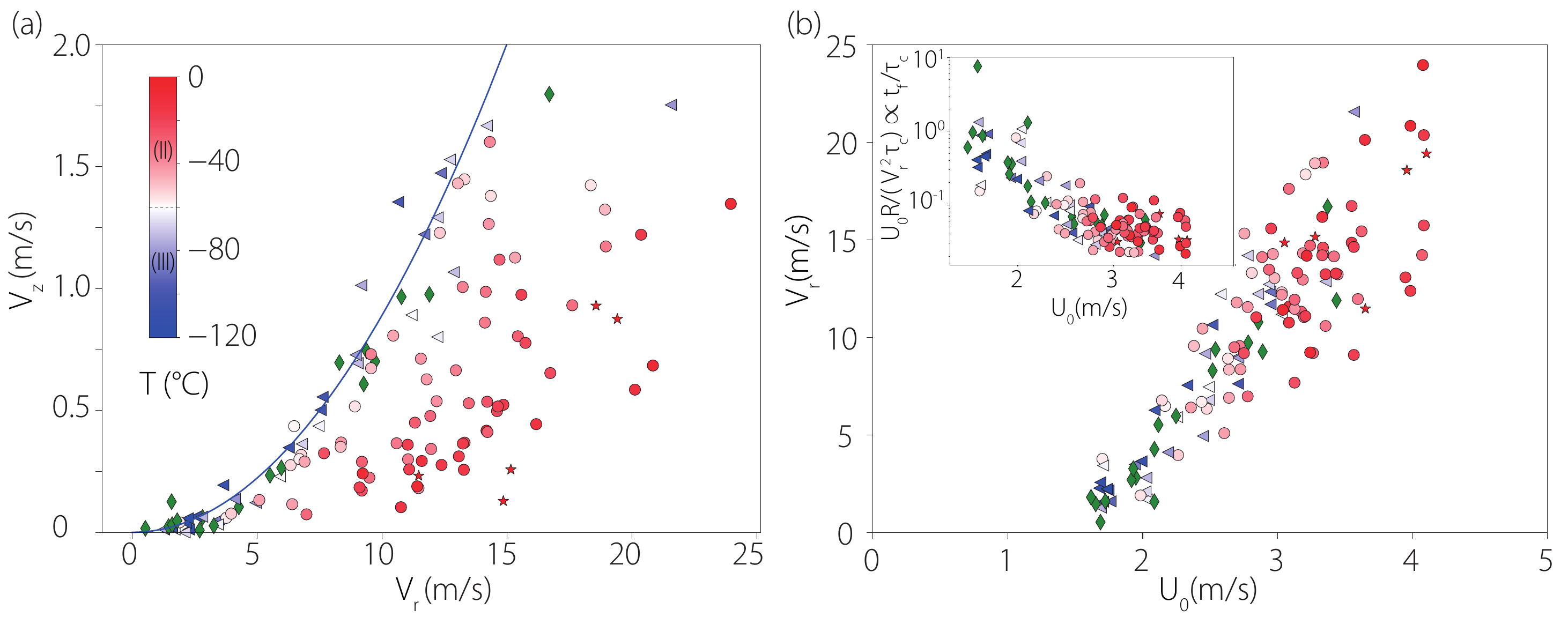}
    \caption{(a) Fastest ejecta vertical velocity as a function of horizontal one: all Region (III) experiments collapse on the superhydrophobic curve, outlined by the blue quadratic curve. (b) Main Figure: fastest ejecta horizontal velocity as a function of the impact velocity. Insert: computed fragmentation time of the corolla. on both figures, no strong dependence on surface temperature is observed. Green symbols for superhydrophobic impacts at ambient temperature, blue-red colors for impacts on silicon wafers at various temperatures ($\blacklozenge$ : Glaco,  {\Large$ \bullet $} : Region (II), $\blacktriangleleft$ : Region (III), {\Large$ \star $} : $T=23^\circ \rm C$ ).  }
    \label{fig:ejecta}
\end{figure}

We now further explore the consequences of that superhydrophobic transition on the morphology of the splashing impacts. We have already emphasised the change in impact aspect with temperature, and the resemblance of splashes in Region (III) with superhydrophobic ones. To characterise this, the radial and vertical velocities of the fastest ejected droplets (respectively denoted by $V_{\rm r}$ and $V_{\rm z}$, as defined on Figure \ref{fig:frise_t} (a)) were measured for each experiment on both sides of the drop for $\rm T<0^\circ C$ and at ambient temperature. 

Figure \ref{fig:ejecta} (a) shows the vertical velocity of the ejecta $V_{\rm z}$ as a function of the horizontal one $V_{\rm r}$. The temperature is colour-coded so that Region (II) corresponds to red dots and Region (III) to blue triangles. The results for the isothermal case (wafer at ambient temperature) are also shown with red stars. The green symbols denote the impacts at ambient temperature on the Glaco coated superhydrophobic surfaces.

Three main findings can be highlighted from these measurements. 
First, the spreading of the points according to their colour shows that for a given value of $V_{\rm r}$ the colder the substrate, the larger the vertical velocity $V_{\rm z}$, resulting in an angle of ejection $\alpha_{\rm e}$ increasing with undercooling. 
Secondly, almost all the blue points (Region (III)) gather around an asymptotic curve outlined by the blue line, demonstrating a saturation effect under this temperature.  
Finally, this asymptotic curve also gathers all the superhydrophobic experiments. 
Together with the observations made in Figure \ref{fig:phases} and \ref{fig:angles}, these results show that cooling down the substrate below $0^\circ \rm C$ is similar to making it more and more hydrophobic, also from the perspective of the ejecta dynamics. Below $-60^\circ \rm C$, the impact systematically looks like one on a superhydrophobic substrate. Interestingly, the saturation curve represented on Figure \ref{fig:ejecta} (a) is well parameterised by an empiric quadratic correlation, $V_{\rm z} \propto V_{\rm r}^2$. This correlation is in contradiction with the few previous attempts of computing the vertical velocity of ejected droplets \cite{burzynski2019role,burzynski2020on-the-splashing} based on a simplification of a more complex model for the ejecta velocity \cite{riboux2015the-diameters}. Note that our data also suggest an influence of the superhydrophobicity on the droplet ejection angle in the isothermal case. This is seen by comparing the star markers at an ambient temperature to the superhydrophobic ambient ones. To our knowledge, it is the first experimental result on this effect and it seems to disagree with a former numerical study \cite{latka2018drop}. %Thess questions would be worth further investigation.
%Insert

Furthermore, we plot the radial velocity of the fastest ejected droplet as a function of the impact velocity U$_{\rm 0}$ on Figure \ref{fig:ejecta} (b). This component of the velocity seems to increase almost linearly with $U_{\rm{0}}$, and most remarkably  no dependence on temperature or wettability seems to appear. From these values, we can evaluate the fragmentation time $t_{\rm f}$ as defined in Figure \ref{fig:frise_t} (a). This evaluation relies on three reasonable assumptions, namely: \textit{(i)} the velocity of the droplet is the one of the corolla at $t_{\rm f}$, \textit{(ii)} the dynamics of the corolla up to $t_{\rm f}$ follows a classical square root of time law for the wetted radius \cite{josserand2003droplet, rioboo2002time}, and \textit{(iii)} these two dynamics do not vary with temperature (in agreement with former studies \cite{gielen2020solidification,kant2020pattern}). The last two assumptions are confirmed by measurements made on all spreading experiments of Regions (II) and (III) (see Appendix \ref{appdynamics}), for which the square root spreading law was observed with a constant prefactor $\pm 10 \%$ and no visible dependence on temperature. In this approach, the fragmentation time can be evaluated as $t_{\rm f} \propto \frac{U_{\rm 0}R}{V_{\rm r}^2}$, which is plotted non-dimensionalised by $\tau_{\rm c}$ on the inset of Figure \ref{fig:ejecta} (b) as a function of $U_{\rm 0}$. The decreasing trend of the fragmentation time as a function of impact velocity is in accordance with previous studies \cite{riboux2015the-diameters}. Again, no dependence on the temperature or the nature of the surface can be observed. Thus, for a given impact velocity, the fragmentation will always take place at the same time if the splash is possible. This unexpected result suggests that solidification does not affect the splash through the fragmentation of the corolla. Meanwhile, the aforementioned increase in ejection angle in Region (II) indicates that the total work applied by the vertical aerodynamic forces during the corolla flight time also increases with substrate undercooling. In the existing models \cite{riboux2015the-diameters,burzynski2019role}, this work only depends on the spreading dynamics, on the air density and on the corolla flight time $t_{\rm f}-t_{\rm to}$. As already mentioned, the spreading dynamics are not affected by temperature. The air density only varies of 30\% over the whole temperature range, which can not explain the relative variations in vertical velocity observed here. We can thus claim that cooling down the substrate increases the corolla flight time $t_{\rm f}-t_{\rm to}$ in Region (II). As $t_{\rm f}$ is independent of $T$, we conclude that solidification favours the early take-off of the corolla in Region (II), up to reaching the asymptotic case of superhydrophobic-like behaviour of Region (III).
This may be possible due to the higher dynamic contact angle reached by the lamella, that can favour an early corolla take-off that is undoubtedly caused by aerodynamic effects over the flowing lamella \cite{zhang2021effect,latka2018drop,latka2017thin-sheet}. 

\section{Conclusion}

To sum up our findings, in this article, we showed that the splashing of water droplets is enhanced on surfaces cooled under the liquid melting temperature. We were able to relate this experimentally to the increase in the lamella dynamic contact angle during spreading, a fact that is thought to be a consequence of ice nucleation within the thermal boundary layer. When the flowing lamella becomes similar to the one flowing over a superhydrophobic surface, the enhancement of splashing is neutralised, so that below a critical temperature the splashing velocity is set to a fixed value, independent of temperature. This behaviour is well captured by an adaptation of existing models. We also investigated the effect of this superhydrophobic transition on the morphology of the splash, by measuring the velocity of the ejecta. Our findings reinforce the idea that regarding drop impact, cooling modifies the substrates dynamic wettability, and suggest an intriguing relationship between this property and the ejection angle of the fragmented droplets. Using classical splashing scenarios, we show that the observed increase in ejection angle with decreasing dynamic wettability can be related to the take-off dynamics of the corolla. Up to now, few studies have looked at these dynamics\cite{latka2018drop,latka2017thin-sheet}, although we show here that it might be an essential ingredient in understanding the ejection features of fragmented droplets both as a function of undercooling and wettability. In further studies, we wish to investigate the latter more systematically in order to provide a better vision of the scenarios discussed here.

\begin{acknowledgments}

We thank the Direction G\'en\'erale de l'Armement (DGA) for financial support and anonymous reviewers for their useful suggestions.

\end{acknowledgments}

\FloatBarrier

\begin{appendix}
\section{Spreading dynamcics of non splashing impacts}\label{appdynamics}

\begin{figure*}[h!]
	\centering
     \begin{subfigure}[b]{0.48\textwidth}
     	\centering
         \includegraphics[width=\textwidth]{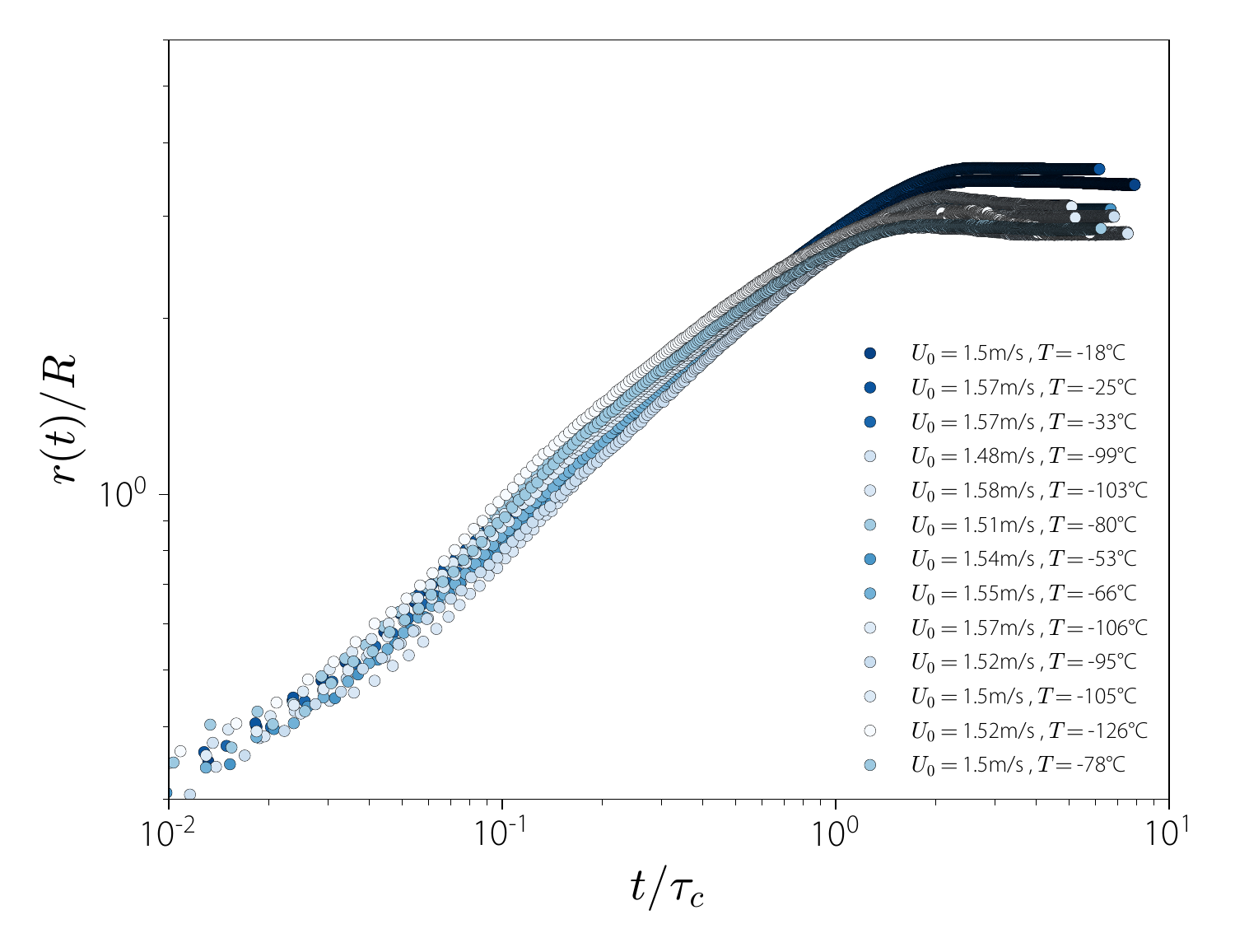}
         \subcaption{$U_{\rm 0} \approx 1.5 \rm{m/s}$}
         \label{fig:cluster15}
     \end{subfigure}
     \hfill
     \begin{subfigure}[b]{0.48\textwidth}
     	\centering
         \includegraphics[width=\textwidth]{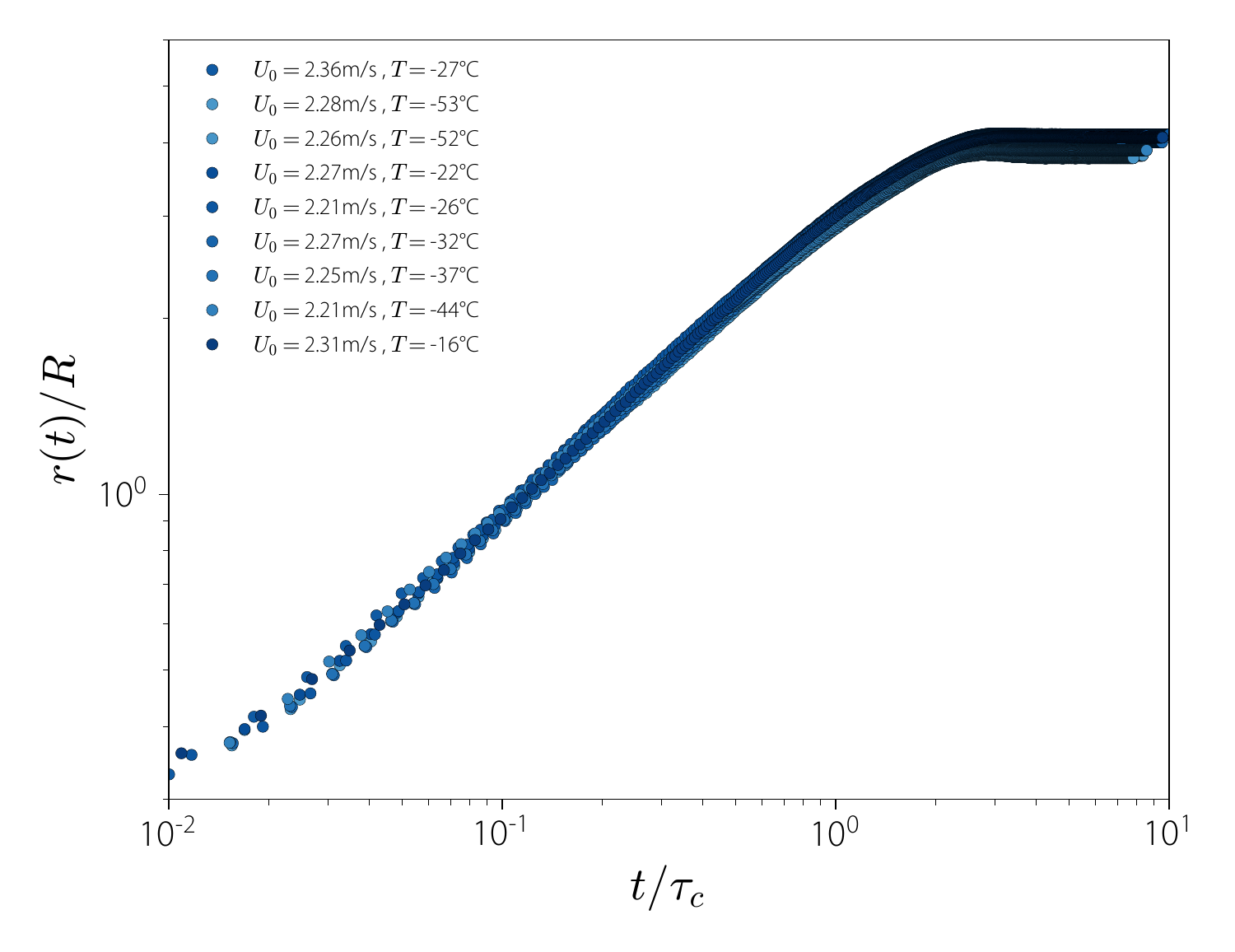}
         \subcaption{$U_{\rm 0} \approx 2.25 \rm{m/s}$}
         \label{fig:cluster22}
     \end{subfigure}

     \caption{Time evolution of the non dimensional wetting radius for two different sets of constant velocity experiments.}
     \label{fig:spreading}
\end{figure*}

Figure \ref{fig:spreading} shows the spreading factor time evolution for two different sets of non-splashing experiments at the same velocity and different temperatures. No clear dependence of the spreading dynamics is visible.

All those dynamics can be fitted to a square root of time law during the first instants: $r(t)=k\sqrt{t}$ \cite{rioboo2002time}. For the position of the root of the lamella, $k_{\rm root}=\sqrt{3U_{\rm 0}R}$ is expected \cite{riboux2014experiments,philippi2016drop}, so we expect $k$ for the tip of the lamella to be slightly larger.

\begin{figure*}[h!]
     \centering
	     \begin{subfigure}[b]{0.48\textwidth}
	         \centering
	         \includegraphics[width=\textwidth]{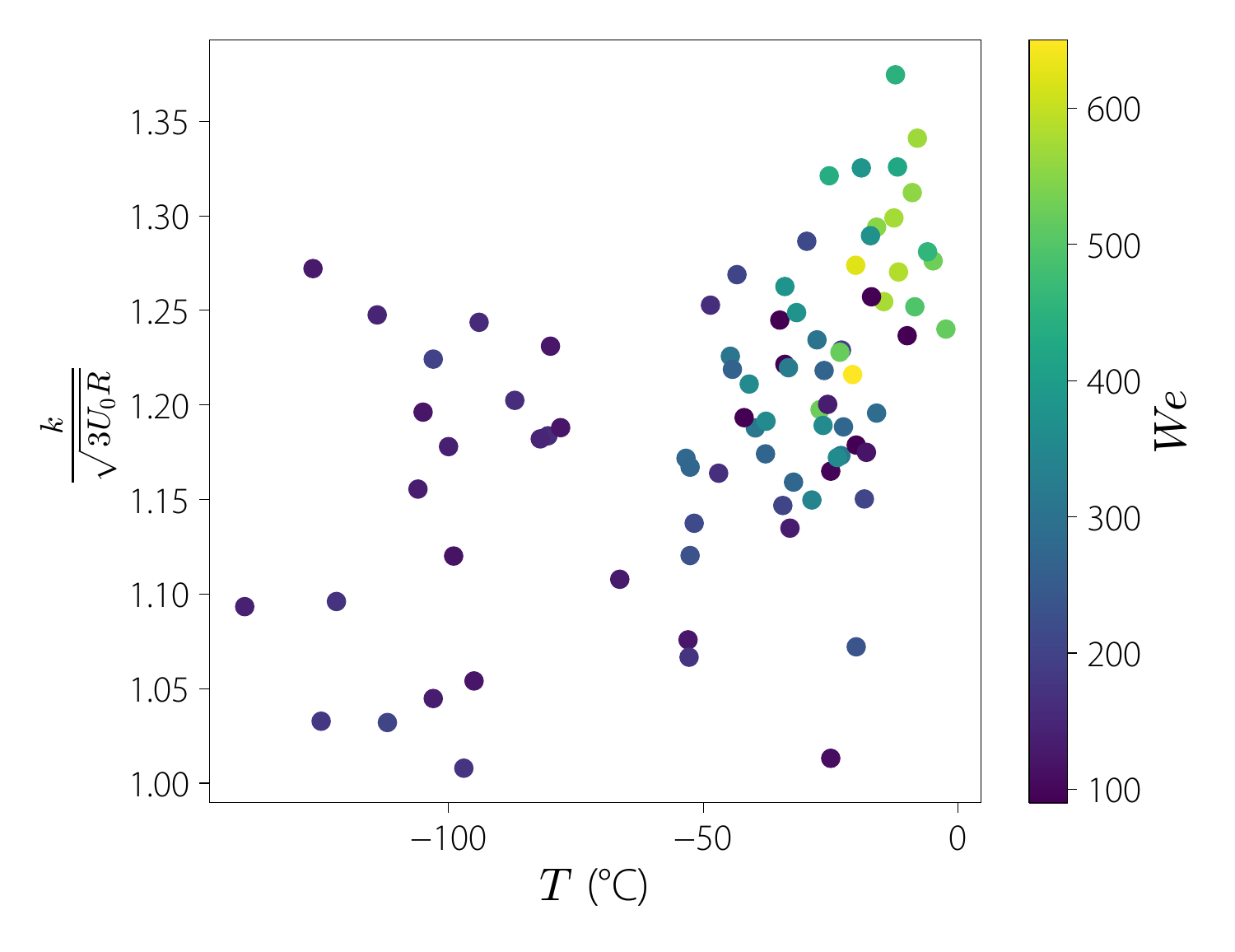}
	         \subcaption{Temperature evolution of the spreading prefactor}
	         \label{fig:kT}
	     \end{subfigure}
	     \hfill
	     \begin{subfigure}[b]{0.48\textwidth}
	         \centering
	         \includegraphics[width=\textwidth]{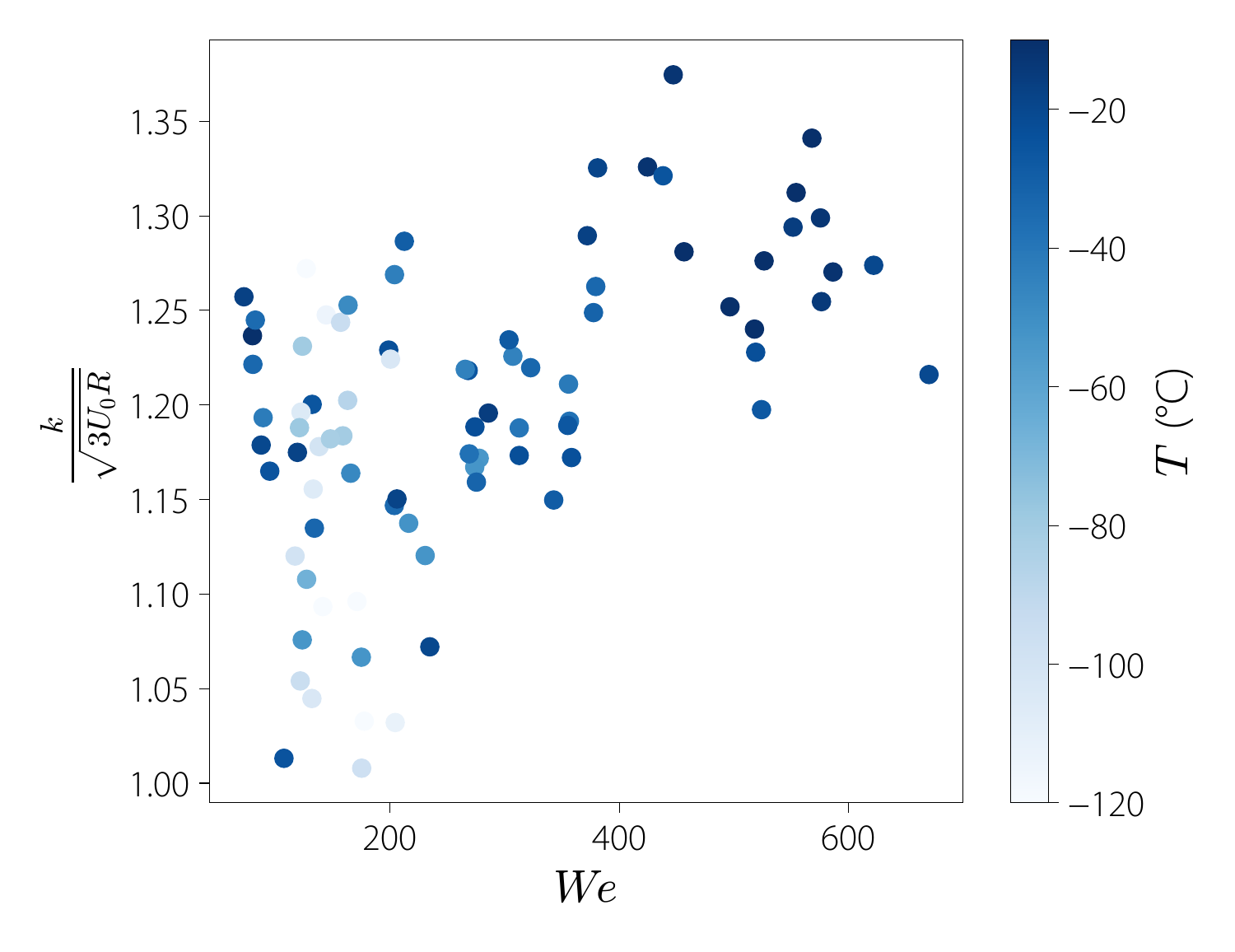}
	         \subcaption{Weber evolution of the spreading prefactor}
	         \label{fig:kWe}
	     \end{subfigure}
     
     \caption{Fitted values of the spreading prefactor for all non splashing experiments of Regions (II) and (III)}
     \label{fig:k}
     
\end{figure*}

On Figure \ref{fig:k}, we plot the prefactors $k$ as a function of surface temperature and impact Weber number. No visible dependence on temperature can be observed, and $k/k_{\rm root}$ is constant $\pm 10\%$, of order unity.

From this spreading law, the spreading lamella velocity is thus $V_{\rm l}(t) \propto 1/\sqrt{t}$, so for a given velocity, $t_{\rm} \propto \frac{U_{\rm 0}R}{V_{\rm l}^2}$.

\FloatBarrier

\section{Dynamic contact angle measurement methodology} \label{appAngle}

As mentioned in the main text, the dynamic contact angle measurement method is based on the one proposed by Quetzeri-Santiago \textit{et al.} \cite{quetzeri-santiago2020on-the-analysis}. 

\begin{figure}
    \centering
    \includegraphics[width=\textwidth]{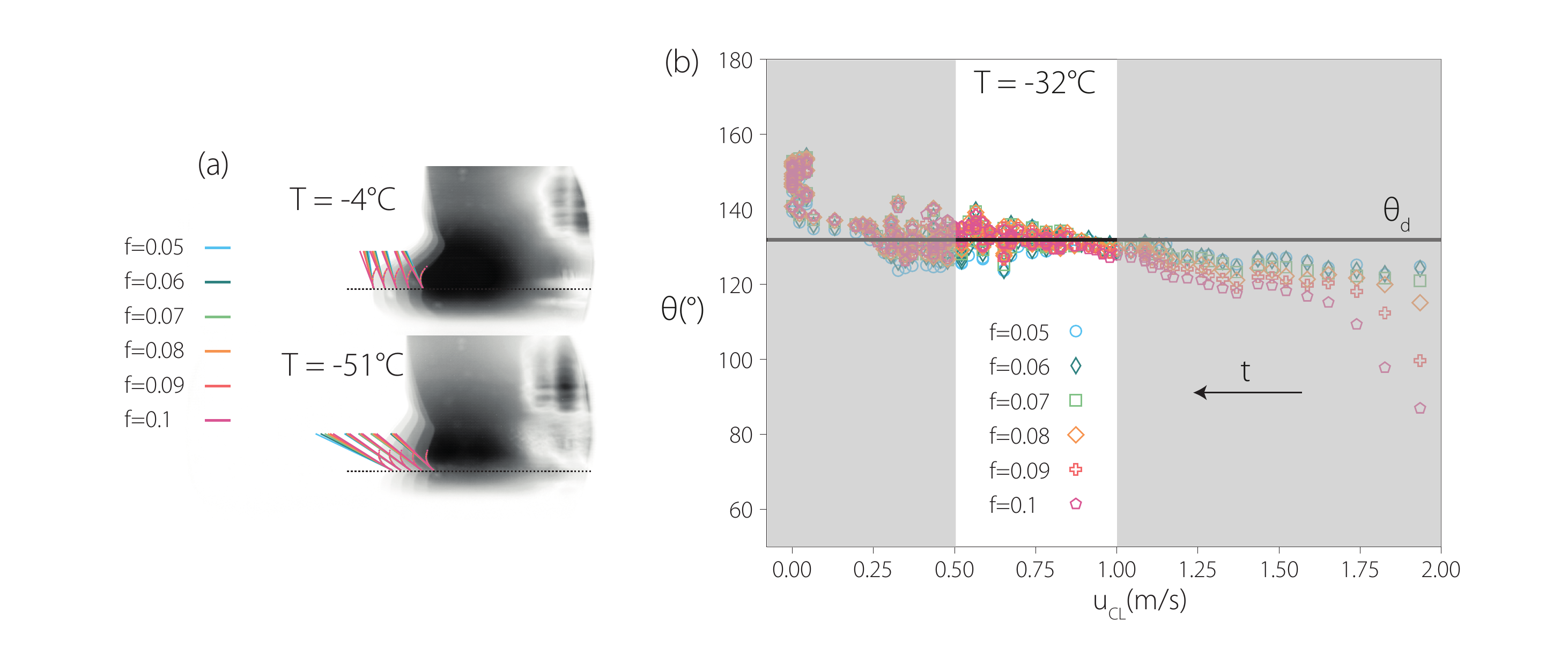}
    \caption{ (a) Chronophotographies of the spreading lamella for drops impacting at $0.7 \rm{m.s^{-1}}$ on surfaces cooled at $-4^\circ \rm C$ and $-51^\circ \rm C$. The colours code for the fraction of drop radius taken to fit the lamella profile. (b) Contact line velocity dependance of the dynamic contact angle during lamella spreading for a drop impacting a wafer at $-37 \rm{^\circ C}$ with a velocity of $U_{\rm 0}=0.7 \rm{m.s^{-1} }$}
    \label{fig:appangle}
\end{figure}

On each frame, we fit a second order polynomial to the $N=f\cdot R_{\rm px}$ first pixels of the lamella, starting from the contact line, where $f$ is a fraction varied between 5 and 10\%, and $R_{\rm px}$ is the drop radius in pixels. For each chosen $f$, the dynamic apparent contact angle is taken as the slope of the tangent at the contact line position. The chronophotographies in Figure \ref{fig:appangle} (a) illustrate the result of that process for different time steps and pixel numbers and two substrate temperatures. From a qualitative perspective, it is clear that the dynamic spreading angle increases as the temperature is decreased. 

The typical contact line velocity dependence of the dynamic angle is shown on Figure \ref{fig:appangle} (b) for an experiment. At high velocities, the results for the different fractions are highly dispersed. At low contact line velocities, the angle increases due to the progressive pinning of the contact line. 
In order to characterise quantitatively each experiment by a unique angle, we removed all time steps where the deviation on the measured angle between the tested $f$ is greater than $2.5^\circ$, and then averaged the angle values over times where the contact line velocity is within the range $[0.5,1]\: \rm{m.s^{-1}}$. The lower bound is chosen in order to avoid the effects related to contact line pinning, the higher one is chosen so that for all experiments, a significant contact angle could be measured at that veocity (ie, an angle for which the deviation between the different fractions was below $2.5^\circ$). Note that it is lower than the one chosen by Quetzeri-Santiago \textit{et al.} \cite{quetzeri-santiago2020on-the-analysis}, $2 \: \rm{m.s^{-1}}$, a velocity that was not in reach of our experimental resolution. Also note that, as the angle increases at low velocities, we do not make use of the denomination "maximal dynamic contact angle" proposed by these authors \cite{quetzeri-santiago2019role}, but rather talk about a dynamic contact angle, which is in fact the one measured at high lamella velocities.

\FloatBarrier

\section{Impact outcomes on cold superhydrophobic surfaces} \label{appSH}

\begin{figure}[h!]
    \centering
    \includegraphics[width=0.49\textwidth]{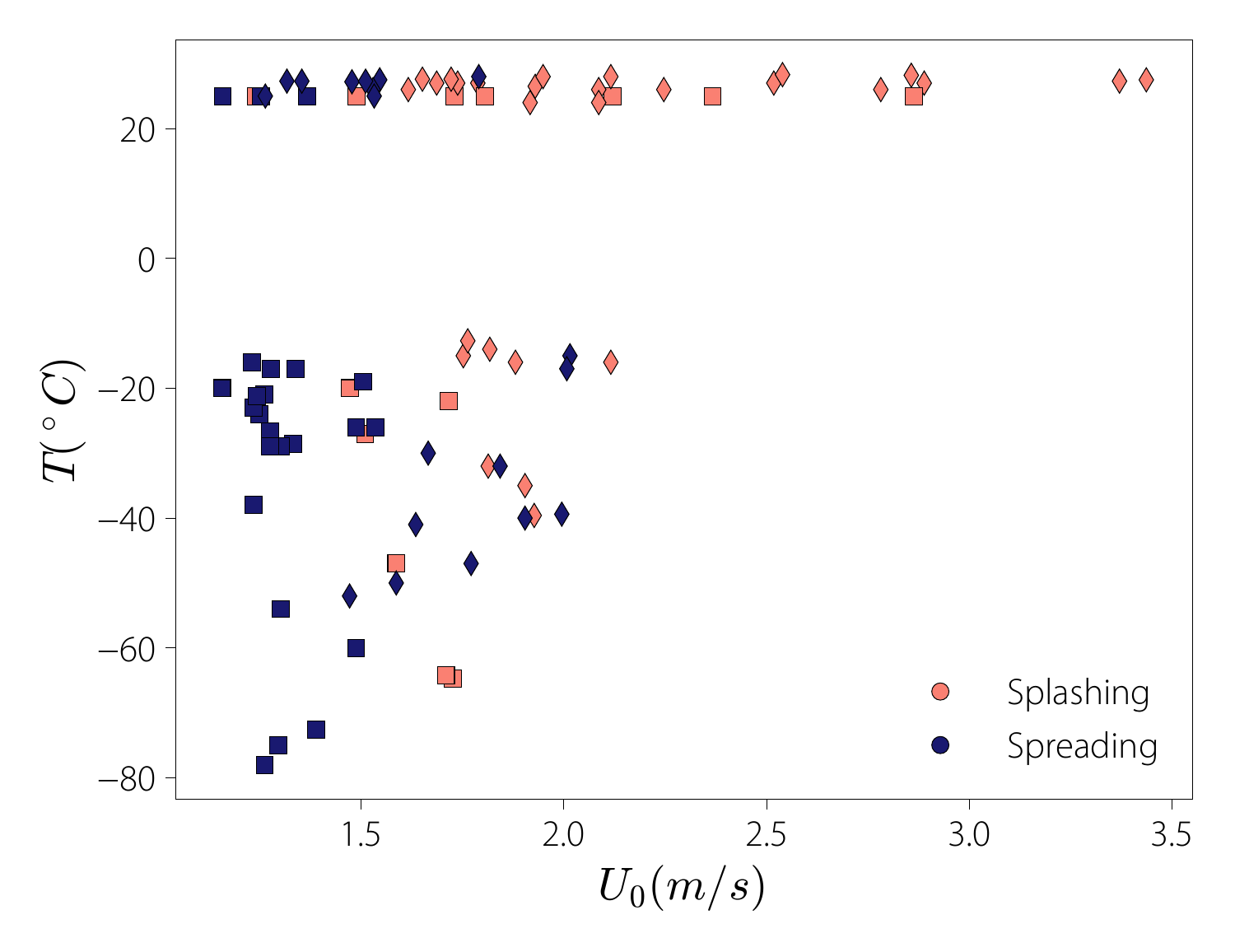}
    \caption{Phase diagram of impact outcome on superhydrophobic surfaces ($\blacklozenge$ : Glaco, $\blacksquare$ : UED). No visible dependence of splashing transition velocity with temperature.}
    \label{fig:sh}
\end{figure}

Figure \ref{fig:sh} shows the phase plot on the Glaco coated superhydrophobic surface, and on a surface coated with commercial Ultra Ever Dry (UED). As this latter surface is harder to characterize and macroscopically less smooth, it was not used in the article. No visible dependence of the splashing transition velocity on temperature is observed.
\FloatBarrier

\FloatBarrier
\section{Impact videos}

Videos of some impacts of Figure \ref{fig:v_constant} are provided in supplementary materials.

\end{appendix}

\FloatBarrier
\bibliography{biblio.bib}% Produces the bibliography via BibTeX.

\end{document}